# Complexity of ITL model checking: some well-behaved fragments of the interval logic HS

Alberto Molinari and Angelo Montanari
*Department of Mathematics and Computer Science*
*University of Udine, Italy*
Email: molinari.alberto@gmail.com; angelo.montanari@uniud.it

Adriano Peron
*Department of Electrical Engineering and Information Technology*
*University of Napoli Federico II, Italy*
Email: adrperon@unina.it

*Abstract*—Model checking has been successfully used in many computer science fields, including artificial intelligence, theoretical computer science, and databases. Most of the proposed solutions make use of classical, point-based temporal logics, while little work has been done in the interval temporal logic setting. Recently, a non-elementary model checking algorithm for Halpern and Shoham's modal logic of time intervals HS over finite Kripke structures (under the homogeneity assumption) and an EXPSPACE model checking procedure for two meaningful fragments of it have been proposed. In this paper, we show that more efficient model checking procedures can be developed for some expressive enough fragments of HS.

*Keywords*-Interval Temporal Logic; Model Checking; Complexity

## I. INTRODUCTION

Model checking algorithms allow one to verify a formal specification of the desired properties of a system against a model of its behaviour. In the standard formulation, systems are described as (finite) labelled state-transition graphs (Kripke structures) and point-based, linear or branching temporal logics (e.g., LTL or CTL) are used to constrain the way in which the truth value of the state-labelling proposition letters changes along the paths of the Kripke structure.

Point-based temporal logics are well-suited for a variety of application domains; however, there are some relevant temporal features, such as actions with duration, accomplishments, and temporal aggregations, that are inherently "interval-based" and thus cannot be expressed by them. Here interval temporal logics (ITLs), that take intervals, instead of points, as their primitive entities, come into play, providing an alternative setting for reasoning about time [8], [23], [24]. To check interval properties of computations, one needs to collect information about states into computation stretches: this amounts to interpret each finite path of a Kripke structure as an interval, and to suitably define its labelling on the basis of the labelling of the states that compose it. Such an increase in expressiveness makes ITLs well suited for a number of applications in the areas of formal verification, computational linguistics, databases, planning, and multi-agent systems, e.g., [2], [7], [18], [20], [25], but, unfortunately, also undecidable.

A prominent position among ITLs is occupied by Halpern and Shoham's modal logic of time intervals, abbreviated HS [8]. HS features one modality for each of the 13 possible ordering relations between pairs of intervals (the so-called Allen's relations [1]), apart from the equality relation. In [8], it has been shown that the *satisfiability problem for HS* interpreted over all relevant (classes of) linear orders is highly undecidable. Since then, a lot of work has been done on satisfiability for HS fragments, which showed that undecidability rules over them [3], [10], [13]. However, meaningful exceptions exist, e.g., the interval logic of temporal neighbourhood $A\overline{A}$ and the logic of sub-intervals D [4]–[6], [17].

In this paper, we focus our attention on the *model checking problem for HS* (not on satisfiability checking), which only very recently entered the research agenda for ITLs.

In [16], Montanari et al. addressed the model checking problem for full HS over finite Kripke structures (under the homogeneity assumption [21]). They introduced the basic elements of the picture, namely, the interpretation of HS formulas over (abstract) interval models, the mapping of finite Kripke structures into (abstract) interval models, and the notion of track descriptor, and they proved a small model theorem showing the non-elementary decidability of the problem. In [14], Molinari et al. gave a lower bound to the complexity of the problem, which is EXPSPACE-hard, if a succinct encoding of formulas is used, PSPACE-hard otherwise. In [15], Molinari et al. showed that model checking for the HS fragment $A\overline{A}B\overline{B}E$ (resp., $A\overline{A}E\overline{E}B$), whose modalities allow one to access intervals which are met by/meet the current one, or are prefixes (resp., suffixes) or right/left-extensions of it, is in EXPSPACE. Moreover, they proved that the problem is NEXP-hard, if a succinct encoding of formulas is used, NP-hard otherwise. Finally, they showed that formulas that satisfy a (constant) bound to the nesting depth of $\langle B \rangle$ (resp., $\langle E \rangle$) modalities can be checked in polynomial working space.

In [11], [12], Lomuscio and Michaliszyn studied the model checking problem for epistemic extensions of some HS fragments. Their semantic assumptions differ from those made in [16], making it difficult to compare the two research lines. In [11], they focused their attention on the fragment BED, whose modalities allow one to respectively access prefixes, suffixes, and sub-intervals of the current interval, extended with epistemic modalities. They considered a restricted form of model checking, which verifies the given specification against a single (finite) initial computation interval (not all possible ones), and proved that it is a PSPACE-complete problem. Moreover, they showed that the problem for the purely temporal fragment of the logic is in PTIME. This last result does not come as a surprise as it trades expressiveness for efficiency: BED modalities allow one to access only sub-intervals of the initial one, whose number is quadratic in the length (number of states) of the initial interval. In [12], they showed that the picture drastically changes with HS fragments that allow one to access infinitely many tracks/intervals. In particular, they proved that this is the case with the fragment A$\overline{B}$L, whose modalities allow one to access intervals which are met by (resp., extend to the right, follow) the current one, extended with epistemic modalities: the problem turns out to be decidable with a non-elementary upper bound.

Table I
ALLEN'S RELATIONS AND CORRESPONDING HS MODALITIES.

| Allen's relation | HS | Definition w.r.t. interval structures | Example |
|---|---|---|---|
| MEETS | $\langle A \rangle$ | $[x,y]\mathcal{R}_A[v,z] \iff y = v$ | |
| BEFORE | $\langle L \rangle$ | $[x,y]\mathcal{R}_L[v,z] \iff y < v$ | |
| STARTED-BY | $\langle B \rangle$ | $[x,y]\mathcal{R}_B[v,z] \iff x = v \wedge z < y$ | |
| FINISHED-BY | $\langle E \rangle$ | $[x,y]\mathcal{R}_E[v,z] \iff y = z \wedge x < v$ | |
| CONTAINS | $\langle D \rangle$ | $[x,y]\mathcal{R}_D[v,z] \iff x < v \wedge z < y$ | |
| OVERLAPS | $\langle O \rangle$ | $[x,y]\mathcal{R}_O[v,z] \iff x < v < y < z$ | |

In this paper, we improve on the results in [15] by identifying some well-behaved HS fragments, which are still expressive enough to capture meaningful interval properties and computationally more efficient. The rest of the paper is organized as follows. In Section II we provide some background knowledge. In Section III we give an overview of the main results of the paper and relate them to known ones. Then, in Section IV we study the existential and universal fragments of A$\overline{A}$BE, while in Section V we analyze the fragments A$\overline{A}$B$\overline{E}$ and A$\overline{B}$. Conclusions provide a short assessment of the work and outline future research directions.

## II. PRELIMINARIES

### A. The interval temporal logic HS

An interval algebra to reason about intervals and their relative order was first proposed by Allen in [1]. A systematic logical study of interval representation and reasoning was then done by Halpern and Shoham, who introduced the interval temporal logic HS featuring one modality for each Allen's relation, except for equality [8]. Table I depicts 6 of the 13 Allen's relations, together with the corresponding HS (existential) modalities. The other 7 relations are the 6 inverse relations (given a binary relation $\mathcal{R}$, the inverse relation $\overline{\mathcal{R}}$ is such that $b\overline{\mathcal{R}}a$ if and only if $a\mathcal{R}b$) and equality.

The language of HS consists of a set of proposition letters $\mathcal{AP}$, the Boolean connectives $\neg$ and $\wedge$, and a temporal modality for each of the (non trivial) Allen's relations, namely, $\langle A \rangle$, $\langle L \rangle$, $\langle B \rangle$, $\langle E \rangle$, $\langle D \rangle$, $\langle O \rangle$, $\langle \overline{A} \rangle$, $\langle \overline{L} \rangle$, $\langle \overline{B} \rangle$, $\langle \overline{E} \rangle$, $\langle \overline{D} \rangle$, and $\langle \overline{O} \rangle$.

HS formulas are defined by the following grammar:

$$\psi ::= p \mid \neg\psi \mid \psi \wedge \psi \mid \langle X \rangle \psi \mid \langle \overline{X} \rangle \psi,$$

where $p \in \mathcal{AP}$ and $X \in \{A, L, B, E, D, O\}$. In the following, we will make use of the standard abbreviations of propositional logic. Furthermore, for all $X$, dual universal modalities $[X]\psi$ and $[\overline{X}]\psi$ are respectively defined as $\neg\langle X \rangle \neg\psi$ and $\neg\langle \overline{X} \rangle \neg\psi$.

Given any subset of Allen's relations $\{X_1, \cdots, X_n\}$, we denote by $X_1 \cdots X_n$ the HS fragment that features modalities $\langle X_1 \rangle, \cdots, \langle X_n \rangle$ only.

We assume the strict semantics of HS: only intervals consisting of at least two points are allowed (no point-intervals)[1]. Under this assumption, all HS modalities can be expressed in terms of modalities $\langle A \rangle$, $\langle B \rangle$, $\langle E \rangle$, $\langle \overline{A} \rangle$, $\langle \overline{B} \rangle$, and $\langle \overline{E} \rangle$ [23]. HS can be viewed as a multi-modal logic with these 6 primitive modalities, and its semantics can be defined over a multi-modal Kripke structure, called here an *abstract interval model*, in which (strict) intervals are treated as atomic objects and Allen's relations as binary relations between pairs of intervals.

*Definition 1 (Abstract interval model [14]):* An *abstract interval model* is a tuple $\mathcal{A} = (\mathcal{AP}, \mathbb{I}, A_\mathbb{I}, B_\mathbb{I}, E_\mathbb{I}, \sigma)$, where:

- $\mathcal{AP}$ is a finite set of proposition letters;
- $\mathbb{I}$ is a possibly infinite set of atomic objects (worlds);
- $A_\mathbb{I}$, $B_\mathbb{I}$, $E_\mathbb{I}$ are three binary relations over $\mathbb{I}$;
- $\sigma : \mathbb{I} \mapsto 2^{\mathcal{AP}}$ is a (total) labeling function, which assigns a set of proposition letters to each world.

In the interval setting, $\mathbb{I}$ is interpreted as a set of intervals and $A_\mathbb{I}$, $B_\mathbb{I}$, and $E_\mathbb{I}$ as Allen's interval relations $A$ (*meets*), $B$ (*started-by*), and $E$ (*finished-by*), resp., and $\sigma$ assigns to each interval the set of proposition letters that hold over it.

---
[1]Strict semantics can be easily "relaxed" to include point-intervals. All results we are going to prove hold for the non-strict semantics as well.

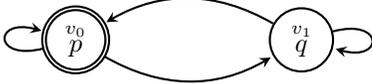

Figure 1. The Kripke structure $\mathcal{K}_{Equiv}$.

Given an abstract interval model $\mathcal{A} = (\mathcal{AP}, \mathbb{I}, A_\mathbb{I}, B_\mathbb{I}, E_\mathbb{I}, \sigma)$ and an interval $I \in \mathbb{I}$, the truth of an HS formula over $I$ is inductively defined as follows:

- $\mathcal{A}, I \models p$ iff $p \in \sigma(I)$, for any $p \in \mathcal{AP}$;
- $\mathcal{A}, I \models \neg \psi$ iff it is not true that $\mathcal{A}, I \models \psi$;
- $\mathcal{A}, I \models \psi \wedge \phi$ iff $\mathcal{A}, I \models \psi$ and $\mathcal{A}, I \models \phi$;
- $\mathcal{A}, I \models \langle X \rangle \psi$, for $X \in \{A, B, E\}$, iff there exists $J \in \mathbb{I}$ such that $I \, X_\mathbb{I} \, J$ and $\mathcal{A}, J \models \psi$;
- $\mathcal{A}, I \models \langle \overline{X} \rangle \psi$, for $\overline{X} \in \{\overline{A}, \overline{B}, \overline{E}\}$, iff there exists $J \in \mathbb{I}$ such that $J \, X_\mathbb{I} \, I$ and $\mathcal{A}, J \models \psi$.

*Satisfiability* and *validity* are defined as usual: an HS formula $\psi$ is satisfiable if there are an interval model $\mathcal{A}$ and a world/interval $I$ such that $\mathcal{A}, I \models \psi$; $\psi$ is valid, denoted $\models \psi$, if $\mathcal{A}, I \models \psi$ for all worlds/intervals $I$ of any interval model $\mathcal{A}$.

*B. Kripke structures and abstract interval models*

Finite state systems are usually modelled as finite Kripke structures. In [16], the authors define a mapping from Kripke structures to abstract interval models that makes it possible to specify properties of systems by means of HS formulas.

*Definition 2:* A *finite Kripke structure* $\mathcal{K}$ is a tuple $(\mathcal{AP}, W, \delta, \mu, w_0)$, where $\mathcal{AP}$ is a set of proposition letters, $W$ is a finite set of states, $\delta \subseteq W \times W$ is a left-total relation between pairs of states, $\mu : W \mapsto 2^{\mathcal{AP}}$ is a total labelling function, and $w_0 \in W$ is the initial state.

For all $w \in W$, $\mu(w)$ is the set of proposition letters that hold at $w$, while $\delta$ is the transition relation that constrains the evolution of the system over time.

*Example 1:* Figure 1 depicts the finite Kripke structure $\mathcal{K}_{Equiv} = (\{p, q\}, \{v_0, v_1\}, \delta, \mu, v_0)$, where $\mu(v_0) = \{p\}$, $\mu(v_1) = \{q\}$, and $\delta = \{(v_0, v_0), (v_0, v_1), (v_1, v_0), (v_1, v_1)\}$ [16]. The initial state $v_0$ is denoted by a double circle.

*Definition 3:* A *track* $\rho$ over a finite Kripke structure $\mathcal{K} = (\mathcal{AP}, W, \delta, \mu, w_0)$ is a finite sequence of states $v_0 \cdots v_n$, with $n \geq 1$, such that for all $i \in \{0, \cdots, n-1\}$, $(v_i, v_{i+1}) \in \delta$.

Let $\mathrm{Trk}_\mathcal{K}$ be the (infinite) set of all tracks over a finite Kripke structure $\mathcal{K}$. For any track $\rho = v_0 \cdots v_n \in \mathrm{Trk}_\mathcal{K}$, we define: $|\rho| = n+1$, $\rho(i) = v_i$, $\mathrm{states}(\rho) = \{v_0, \cdots, v_n\} \subseteq W$, $\mathrm{intstates}(\rho) = \{v_1, \cdots, v_{n-1}\} \subseteq W$, $\mathrm{fst}(\rho) = v_0$ and $\mathrm{lst}(\rho) = v_n$. If $\mathrm{fst}(\rho) = w_0$, $\rho$ is called an *initial track*. Let $\rho(i,j) = v_i \cdots v_j$, for $0 \leq i < j \leq |\rho|-1$, be a subtrack of $\rho$. $\mathrm{Pref}(\rho) = \{\rho(0,i) \mid 1 \leq i \leq |\rho|-2\}$ and $\mathrm{Suff}(\rho) = \{\rho(i, |\rho|-1) \mid 1 \leq i \leq |\rho|-2\}$ are the sets of all proper prefixes and suffixes of $\rho$, respectively. Notice that the length of tracks, prefixes, and suffixes is greater than 1, as they will be mapped into strict intervals. Finally by $\rho \cdot \rho'$ we denote the concatenation of the tracks $\rho$ and $\rho'$.

An abstract interval model (over $\mathrm{Trk}_\mathcal{K}$) can be naturally associated with a finite Kripke structure by interpreting every track as an interval bounded by its first and last states [14].

*Definition 4:* The *abstract interval model induced by a finite Kripke structure* $\mathcal{K} = (\mathcal{AP}, W, \delta, \mu, w_0)$ is $\mathcal{A}_\mathcal{K} = (\mathcal{AP}, \mathbb{I}, A_\mathbb{I}, B_\mathbb{I}, E_\mathbb{I}, \sigma)$, where:

- $\mathbb{I} = \mathrm{Trk}_\mathcal{K}$,
- $A_\mathbb{I} = \{(\rho, \rho') \in \mathbb{I} \times \mathbb{I} \mid \mathrm{lst}(\rho) = \mathrm{fst}(\rho')\}$,
- $B_\mathbb{I} = \{(\rho, \rho') \in \mathbb{I} \times \mathbb{I} \mid \rho' \in \mathrm{Pref}(\rho)\}$,
- $E_\mathbb{I} = \{(\rho, \rho') \in \mathbb{I} \times \mathbb{I} \mid \rho' \in \mathrm{Suff}(\rho)\}$,
- $\sigma : \mathbb{I} \mapsto 2^{\mathcal{AP}}$ is such that $\sigma(\rho) = \bigcap_{w \in \mathrm{states}(\rho)} \mu(w)$, for all $\rho \in \mathbb{I}$.

Relations $A_\mathbb{I}$, $B_\mathbb{I}$, and $E_\mathbb{I}$ are interpreted as Allen's relations $A$, $B$, and $E$, respectively. Moreover, according to the definition of $\sigma$, $p \in \mathcal{AP}$ holds over $\rho = v_0 \cdots v_n$ iff it holds over all the states $v_0, \cdots, v_n$ of $\rho$. This conforms to the *homogeneity principle*, according to which a proposition letter holds over an interval iff it holds over all of its subintervals.

Since $\mathcal{K}$ has loops ($\delta$ is left-total), the number of tracks of $\mathcal{K}$, and thus the number of intervals of $\mathcal{A}_\mathcal{K}$, is infinite.

Satisfiability of an HS formula over a finite Kripke structure can be given in terms of induced abstract interval models.

*Definition 5:* Let $\mathcal{K}$ be a finite Kripke structure, $\rho$ be a track in $\mathrm{Trk}_\mathcal{K}$, and $\psi$ be an HS formula. We say that the pair $(\mathcal{K}, \rho)$ satisfies $\psi$, denoted by $\mathcal{K}, \rho \models \psi$, iff it holds that $\mathcal{A}_\mathcal{K}, \rho \models \psi$.

The *model checking problem* for HS over finite Kripke structures is the problem of deciding whether $\mathcal{K} \models \psi$.

*Definition 6:* Let $\mathcal{K}$ be a finite Kripke structure and $\psi$ be an HS formula. We say that $\mathcal{K}$ models $\psi$, denoted by $\mathcal{K} \models \psi$, iff for all *initial* tracks $\rho \in \mathrm{Trk}_\mathcal{K}$ it holds that $\mathcal{K}, \rho \models \psi$.

Some examples of meaningful properties of tracks that can be expressed in HS can be found in [14].

III. THE GENERAL PICTURE

In [14], Molinari et al. showed that, given a finite Kripke structure $\mathcal{K}$ and a bound $k$ on the structural complexity of HS formulas (that is, on the nesting depth of $E$ and $B$ modalities), it is possible to obtain a *finite* representation for $\mathcal{A}_\mathcal{K}$, which is equivalent to $\mathcal{A}_\mathcal{K}$ with respect to satisfiability of HS formulas with structural complexity less than or equal to $k$. Then, by exploiting such a representation, they proved that the model checking problem for (full) HS is decidable (the given algorithm has a non-elementary upper bound). Moreover, they showed that the problem for the fragment $A\overline{A}BE$, and thus for full HS, is PSPACE-hard (EXPSPACE-hard if a suitable succinct encoding of formulas is exploited).

In [15], Molinari et al. devised an EXPSPACE model checking algorithm for the fragments $A\overline{A}B\overline{B}E$ and $A\overline{A}E\overline{B}E$, that needs to consider only a subset of relatively short tracks: for any given bound $k$ on the complexity of formulas, they defined an equivalence relation over tracks of finite index, and they showed that model checking can be restricted to track representatives of bounded length. In addition, they proved that the problem is NP-hard (if a suitable succinct encoding of formulas is exploited, the algorithm remains in EXPSPACE, but a NEXPTIME lower bound can be given).

Here, we identify some well-behaved HS fragments, namely, $\forall A\overline{A}BE$ (and $\exists A\overline{A}BE$), $A\overline{A}BE$, and $A\overline{A}$, which are still expressive enough to capture meaningful interval properties of state-transition systems and whose model checking problem exhibits a considerably lower computational complexity. The simple example below shows some of these fragments at work.

*Example 2:* Let $\mathcal{K} = (\mathcal{AP}, W, \delta, \mu, w_0)$, with $\mathcal{AP} = \{r_0, r_1, e_0, e_1, x_0\}$, be the Kripke structure of Figure 2, that models the interactions between a scheduler $\mathcal{S}$ and two processes, $\mathcal{P}_0$ and $\mathcal{P}_1$, which possibly ask for a shared resource. At the initial state $w_0$, $\mathcal{S}$ has not received any request from the processes yet, while in $w_1$ (resp., $w_2$) only $\mathcal{P}_0$ (resp., $\mathcal{P}_1$) has sent a request, and thus $r_0$ (resp., $r_1$) holds. As long as at most one process has sent a request, $\mathcal{S}$ is not forced to allocate the resource ($w_1$ and $w_2$ have self loops). At $w_3$, both $\mathcal{P}_0$ and $\mathcal{P}_1$ are waiting for the shared resource, and hence both $r_0$ and $r_1$ hold there. State $w_3$ has transitions only towards $w_4$, $w_6$, and $w_8$. At $w_4$ (resp., $w_6$) $\mathcal{P}_1$ (resp., $\mathcal{P}_0$) can access the resource: $e_1$ (resp., $e_0$) holds in $w_4 w_5$ (resp., $w_6 w_7$). However, a faulty transition may be taken from $w_3$: in $w_8$ and $w_9$ both $\mathcal{P}_0$ and $\mathcal{P}_1$ are using the resource (both $e_0$ and $e_1$ hold in $w_8 w_9$). Finally, from $w_5$, $w_7$, and $w_9$ the system can only move to $w_0$, where $\mathcal{S}$ waits for new requests from $\mathcal{P}_0$ and $\mathcal{P}_1$.

Now, let $\mathcal{P}$ be the set $\{r_0, r_1, e_0, e_1\}$ and let $x_0$ be an auxiliary proposition letter labelling the states $w_0$, $w_1$, $w_6$, and $w_7$, where $\mathcal{S}$ and $\mathcal{P}_0$, but not $\mathcal{P}_1$, are active.

It holds that $\mathcal{K} \models [A]\psi$ (equivalently, $\mathcal{K} \models [E]\psi$) iff $\psi$ holds over any (reachable) computation sub-interval.

It can also be checked that $\mathcal{K} \not\models [E]\neg(e_0 \wedge e_1)$ (this formula is in $\forall A\overline{A}BE$), i.e., mutual exclusion is not guaranteed, as the faulty transition leading to $w_8$ may be taken at $w_3$, and then both $\mathcal{P}_0$ and $\mathcal{P}_1$ access the resource in $w_8 w_9$ ($e_0 \wedge e_1$ holds).

On the contrary, it holds that $\mathcal{K} \models [A]\bigl(r_0 \rightarrow \langle A \rangle e_0 \vee \langle A \rangle \langle A \rangle e_0\bigr)$ (in $A\overline{A}$ and $A\overline{A}BE$). Such a formula expresses the following reachability property: if $r_0$ holds over some interval, then there is always a way to reach an interval over which $e_0$ holds. Obviously, this does not mean that all possible computations will necessarily lead to such an interval; however, the system will *never* fall in a state from which it is no more possible to satisfy requests from $\mathcal{P}_0$.

It also holds that $\mathcal{K} \models [A]\bigl(r_0 \wedge r_1 \rightarrow [A](e_0 \vee e_1 \vee \bigwedge_{p \in \mathcal{P}} \neg p)\bigr)$ (in $A\overline{A}$ and $A\overline{A}BE$). Indeed, if both processes send a request to $\mathcal{S}$ (state $w_3$), then it immediately allocates the resource. Formally, if $r_0 \wedge r_1$ holds over some tracks (the only possible cases are $w_3 w_4$, $w_3 w_6$, and $w_3 w_8$), then in any possible subsequent interval of length 2 $e_0 \vee e_1$ holds, that is, $\mathcal{P}_0$ or $\mathcal{P}_1$ are executed, or $\bigwedge_{p \in \mathcal{P}} \neg p$ holds, if we consider tracks longer than 2. On the contrary, if only one process asks for the resource, then $\mathcal{S}$ can arbitrarily delay its allocation, that is, $\mathcal{K} \not\models [A]\bigl(r_0 \rightarrow [A](e_0 \vee \bigwedge_{p \in \mathcal{P}} \neg p)\bigr)$.

Finally, it holds that $\mathcal{K} \models x_0 \rightarrow \langle \overline{B} \rangle x_0$ (in $A\overline{A}BE$), that is, any initial track satisfying $x_0$ (any such track involves states $w_0$, $w_1$, $w_6$, and $w_7$ only) can be extended to the right in such a way that the resulting track still satisfies $x_0$. This amounts to say that there exists a computation in which $\mathcal{P}_1$ starves. Notice that $\mathcal{S}$ and $\mathcal{P}_0$ can continuously interact without waiting for $\mathcal{P}_1$. This is the case, for instance, when $\mathcal{P}_1$ does not ask for the shared resource at all.

In Figure 3, we summarize known (white boxes) and new (grey boxes) results about complexity of model checking for HS fragments. The new results are presented in the next two sections. In Section IV, we deal with the fragment $\forall A\overline{A}BE$, including formulas of $A\overline{A}BE$ in which only universal modalities are allowed and negation can be applied to propositional formulas only. We first provide a coNP model checking algorithm for $\forall A\overline{A}BE$, and then we show that the model checking problem for the pure propositional fragment Prop is coNP-hard. The two results allow us to conclude that the model checking problem for both Prop and $\forall A\overline{A}BE$ is coNP-complete. In addition, upper and lower bounds to the complexity of the problem for $A\overline{A}$ (the logic of temporal neighbourhood) directly follow. In [15], the authors show that the EXPSPACE model checking algorithm for $A\overline{A}B\overline{B}E$ can be suitably tailored to check formulas with a constant nesting depth of $\langle B \rangle$ modalities in polynomial space. Thus, as a particular case, the model checking problem for $A\overline{A}BE$ is in PSPACE. Since $A\overline{A}$ is a fragment of $A\overline{A}BE$ and Prop is a fragment of $A\overline{A}$, complexity of model checking for $A\overline{A}$ is in between coNP and PSPACE. In Section V, we focus our attention on $A\overline{A}BE$ and we prove that the model checking problem for $A\overline{B}$ is PSPACE-hard. PSPACE-completeness of $A\overline{A}BE$ (and $A\overline{B}$) immediately follows. From this, we get for free a strengthening of the lower bound to the complexity of the model checking problem for $A\overline{A}B\overline{B}E$ (in the non-succinct case), which turns out to be PSPACE-hard.

## IV. THE FRAGMENTS $\forall A\overline{A}BE$, $A\overline{A}$, AND Prop

In this section, we first deal with the universal and existential fragments of $A\overline{A}BE$, respectively denoted by $\forall A\overline{A}BE$ and $\exists A\overline{A}BE$, whose formulas are defined as follows:

$$\psi ::= \beta \mid \psi \wedge \psi \mid [A]\psi \mid [B]\psi \mid [E]\psi \mid [\overline{A}]\psi$$

(resp., $\psi ::= \beta \mid \psi \vee \psi \mid \langle A \rangle \psi \mid \langle B \rangle \psi \mid \langle E \rangle \psi \mid \langle \overline{A} \rangle \psi$),

where $\beta ::= p \mid \beta \vee \beta \mid \beta \wedge \beta \mid \neg \beta \mid \bot \mid \top$ with $p \in \mathcal{AP}$.

The intersection of $\forall A\overline{A}BE$ and $\exists A\overline{A}BE$ is the set of pure propositional formulas (Prop). Negations occur in pure propositional formulas only, and formulas with modalities can be combined only by conjunctions (in $\forall A\bar{A}BE$) or disjunctions (in $\exists A\bar{A}BE$). The negation of any $\forall A\bar{A}BE$ formula can be transformed into an equivalent $\exists A\bar{A}BE$ formula (of at most double length), and vice versa, by using De Morgan's laws and the equivalences $[X]\psi \equiv \neg\langle X\rangle\neg\psi$ and $\neg\neg\psi \equiv \psi$.

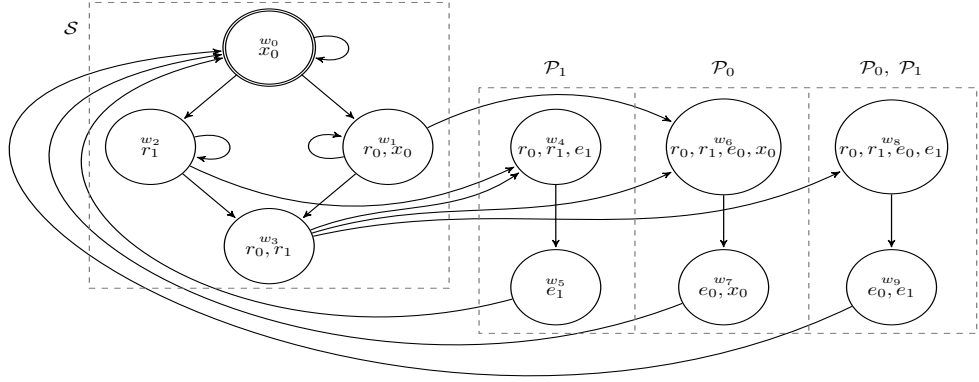

Figure 2. A simple state-transition system.

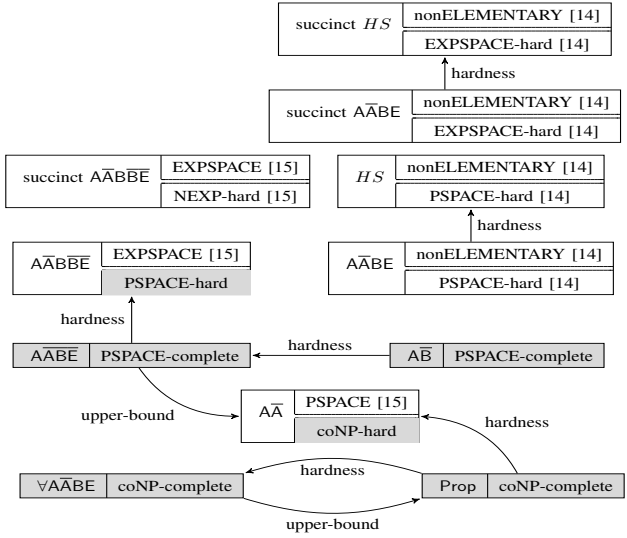

Figure 3. Complexity of model checking for HS fragments.

We now outline a *non-deterministic* algorithm to decide the model checking problem for a $\forall A\bar{A}BE$ formula $\psi$. As usual, the algorithm searches for a counterexample to $\psi$. As we already pointed out, $\neg\psi$ is equivalent to a suitable formula $\psi'$ of the dual fragment $\exists A\bar{A}BE$. Hence, the algorithm looks for an initial track of the finite Kripke structure that satisfies $\psi'$.

For the satisfiability check, we apply the non-deterministic procedure Check (Algorithm 1). Such a procedure uses an abstract representation of tracks called *descriptor element*.

A descriptor element for a finite Kripke structure $\mathcal{K} = (\mathcal{AP}, W, \delta, \mu, w_0)$ is a triple belonging to $W \times 2^W \times W$. The descriptor element for a track $\rho \in \text{Trk}_\mathcal{K}$ is the triple $d = (v_{in}, S, v_{fin})$, where $v_{in} = \text{fst}(\rho)$, $S = \text{intstates}(\rho)$, $v_{fin} = \text{lst}(\rho)$ (we also say that $\rho$ is associated with $d$). The idea is that, to represent a track, we can restrict our attention to the first state, the last state, and the set of states occurring in between, ignoring information about the ordering and multiplicity of their occurrences.

The notion of descriptor element bears analogies with an abstraction technique for discrete time Duration Calculus proposed by Hansen et al. in [9], which on its turn is connected to Parikh images [19] (the notion of descriptor element can be seen as a qualitative analogue of this).

We say that a descriptor element $d$ is *witnessed* (in $\mathcal{K}$) if there exists a track $\rho \in \text{Trk}_\mathcal{K}$ such that $d$ is the descriptor element for $\rho$. Instead of considering tracks of $\mathcal{K}$, which are infinitely many, the model checking algorithm "enumerates" the descriptor elements witnessed in $\mathcal{K}$, which are finitely many. In [15], we proved that if a descriptor element $d$ is witnessed, then there exists a track of length at most $2+|W|^2$ associated with it, and thus the above enumeration involves a non-deterministic polynomial time computation: to generate a (all) witnessed descriptor element(s) with initial state $v$, we just need to non-deterministically visit the unravelling of $\mathcal{K}$ from $v$ up to depth $2 + |W|^2$.

The procedure Check (Algorithm 1) takes as input a Kripke structure $\mathcal{K}$, a formula $\psi$ of $\exists A\bar{A}BE$, and a witnessed descriptor element $d = (v_{in}, S, v_{fin})$ and it returns **Yes** if and only if there exists a track $\rho \in \text{Trk}_\mathcal{K}$ associated with $d$ such that $\mathcal{K}, \rho \models \psi$. The procedure is recursively defined as follows.

If it is called on a Boolean combination $\beta$ of proposition letters (base of the recursion), $VAL(\beta, d)$ evaluates $\beta$ over $d$ in the standard way. The evaluation can be performed in deterministic polynomial time, and if $VAL(\beta, d)$ returns $\top$, then there exists a track associated with $d$ (of length at most quadratic in $|W|$) that satisfies $\beta$.

If $\psi = \psi' \vee \psi''$, where $\psi'$ or $\psi''$ feature some temporal modality, the procedure non-deterministically calls itself on $\psi'$ or $\psi''$ (the construct **Either** $c_1$ **Or** $c_2$ **EndOr** denotes a non-deterministic choice between commands $c_1$ and $c_2$).

**Algorithm 1** Check($\mathcal{K}, \psi, (v_{in}, S, v_{fin})$)

  **if** $\psi = \beta$ **then**            ◁ $\beta$ *is a Boolean combination of propositions*
    **if** $VAL(\beta, (v_{in}, S, v_{fin})) = \top$ **then**
      Yes **else** No
  **else if** $\psi = \varphi_1 \vee \varphi_2$ **then**
    **Either**
      **return** Check($\mathcal{K}, \varphi_1, (v_{in}, S, v_{fin})$)
    **Or**
      **return** Check($\mathcal{K}, \varphi_2, (v_{in}, S, v_{fin})$)
    **EndOr**
  **else if** $\psi = \langle A \rangle\, \varphi$ **then**
    $(v_{fin}, S', v'_{fin}) \leftarrow$ aDescrEl($\mathcal{K}, v_{fin}, $FORW)
    **return** Check($\mathcal{K}, \varphi, (v_{fin}, S', v'_{fin})$)
  **else if** $\psi = \langle \overline{A} \rangle\, \varphi$ **then**
    $(v'_{in}, S', v_{in}) \leftarrow$ aDescrEl($\mathcal{K}, v_{in}, $BACKW)
    **return** Check($\mathcal{K}, \varphi, (v'_{in}, S', v_{in})$)
  **else if** $\psi = \langle B \rangle\, \varphi$ **then**
    $(v'_{in}, S', v'_{fin}) \leftarrow$ aDescrEl($\mathcal{K}, v_{in}, $FORW)    ◁ $v'_{in} = v_{in}$
    **Either**
      **if** $(v'_{in}, S' \cup \{v'_{fin}\}, v_{fin}) = (v_{in}, S, v_{fin})$ and $(v'_{fin}, v_{fin})$ is an edge of $\mathcal{K}$ **then**
        **return** Check($\mathcal{K}, \varphi, (v'_{in}, S', v'_{fin})$)
      **else**
        No
    **Or**
      $(v''_{in}, S'', v''_{fin}) \leftarrow$ aDescrEl($\mathcal{K}, v''_{in}, $FORW),
      where $(v'_{fin}, v''_{in})$ is an edge of $\mathcal{K}$ non-deterministically chosen
      **if** concat$\big((v'_{in}, S', v'_{fin}), (v''_{in}, S'', v''_{fin})\big) = (v_{in}, S, v_{fin})$ **then**
        **return** Check($\mathcal{K}, \varphi, (v'_{in}, S', v'_{fin})$)
      **else**
        No
    **EndOr**
  **else if** $\psi = \langle E \rangle\, \varphi$ **then**
    Symmetric to $\psi = \langle B \rangle\, \varphi$

If $\psi = \langle A \rangle\, \psi'$ (respectively, $\langle \overline{A} \rangle\, \psi'$), the procedure looks for a new descriptor element for a track starting from the final state (respectively, leading to the initial state) of the current descriptor element $d$. To this aim, we use the procedure aDescrEl($\mathcal{K}, v, $FORW) (resp., aDescrEl($\mathcal{K}, v, $BACKW)) which non-deterministically returns a descriptor element $(v'_{in}, S', v'_{fin})$, with $v'_{in} = v$ (resp., $v'_{fin} = v$), witnessed in $\mathcal{K}$ by exploring forward (resp., backward) the unravelling of $\mathcal{K}$ from $v'_{in}$ (resp., from $v'_{fin})^2$. Its complexity is polynomial in $|W|$, since it needs to examine the unravelling of $\mathcal{K}$ from $v$ up to depth $2 + |W|^2$.

If $\psi = \langle B \rangle\, \psi'$, the procedure looks for a new descriptor element $d_1$ and eventually calls itself on $\psi'$ and $d_1$ only if the current descriptor element $d$ results from the "concatenation" of $d_1$ with a suitable descriptor element $d_2$: if $d_1 = (v'_{in}, S', v'_{fin})$ and $d_2 = (v''_{in}, S'', v''_{fin})$, then concat$(d_1, d_2)$ returns $(v'_{in}, S' \cup \{v'_{fin}, v''_{in}\} \cup S'', v''_{fin})$. Notice that if $\rho_1$ and $\rho_2$ are tracks associated with $d_1$ and $d_2$, respectively, then $\rho_1 \cdot \rho_2$ is associated with concat$(d_1, d_2)$.

The following theorem proves soundness and completeness of the Check procedure.

---

[2] By forward (respectively, backward) unravelling of a Kripke structure $\mathcal{K} = (\mathcal{AP}, W, \delta, \mu, w_0)$ from $v \in W$, we refer to the unravelling of the graph $(W, \delta)$ (respectively, $(W, \overline{\delta})$, where $\overline{\delta}$ is the inverse of $\delta$) from $v$.

*Theorem 1:* For any $\exists A\overline{A}BE$ formula $\psi$ and any witnessed descriptor element $d = (v_{in}, S, v_{fin})$, the procedure Check($\mathcal{K}, \psi, d$) has a successful computation iff there exists a track $\rho$ associated with $d$ such that $\mathcal{K}, \rho \models \psi$.

*Proof:* (Soundness) The proof is by induction on the structure of the formula $\psi$.

- $\psi$ is a Boolean combination of propositions $\beta$: let $\rho$ be a witness track for $d$; if check($\mathcal{K}, \beta, d$) has a successful computation, then $VAL(\beta, d)$ is true and so $\mathcal{K}, \rho \models \psi$.
- $\psi = \varphi_1 \vee \varphi_2$: if check($\mathcal{K}, \psi, d$) has a successful computation, then, for some $i \in \{1, 2\}$, check($\mathcal{K}, \varphi_i, d$) has a successful computation. By the inductive hypothesis, there exists $\rho \in \text{Trk}_\mathcal{K}$ associated with $d$ such that $\mathcal{K}, \rho \models \varphi_i$, and thus $\mathcal{K}, \rho \models \varphi_1 \vee \varphi_2$.
- $\psi = \langle A \rangle\, \varphi$: if check($\mathcal{K}, \psi, d$) has a successful computation, then there exists a witnessed $d' = (v'_{in}, S', v'_{fin})$, with $v'_{in} = v_{fin}$, such that check($\mathcal{K}, \varphi, d'$) has a successful computation. By the inductive hypothesis, there exists a track $\rho'$, associated with $d'$, such that $\mathcal{K}, \rho' \models \varphi$. If $\rho$ is a track associated with $d$ (which is witnessed by hypothesis), we have that $\text{lst}(\rho) = \text{fst}(\rho') = v_{fin}$ and, by definition, $\mathcal{K}, \rho \models \psi$.
- $\psi = \langle B \rangle\, \varphi$: if check($\mathcal{K}, \psi, d$) has a successful computation, then we must distinguish two possible cases. (i) There exists $d' = (v_{in}, S', v'_{fin})$, witnessed by a track with $(v'_{fin}, v_{fin}) \in \delta$, such that $(v_{in}, S' \cup \{v'_{fin}\}, v_{fin}) = d$, and check($\mathcal{K}, \varphi, d'$) has a successful computation. By the inductive hypothesis, there exists a track $\rho'$, associated with $d'$, such that $\mathcal{K}, \rho' \models \varphi$. Hence $\mathcal{K}, \rho' \cdot v_{fin} \models \psi$ and $\rho' \cdot v_{fin}$ is associated with $d$.
(ii) There exist $d' = (v_{in}, S', v'_{fin})$, witnessed by a track, and $d'' = (v''_{in}, S'', v''_{fin})$, witnessed by a track as well, such that $(v'_{fin}, v''_{in}) \in \delta$, concat$(d', d'') = d$, and check($\mathcal{K}, \varphi, d'$) has a successful computation. By the inductive hypothesis, there exists a track $\rho'$, associated with $d'$, such that $\mathcal{K}, \rho' \models \varphi$. Hence $\mathcal{K}, \rho' \cdot \rho'' \models \psi$, where $\rho''$ is any track associated with $d''$ and $\rho' \cdot \rho''$ is associated with $d$.

The case $\psi = \langle \overline{A} \rangle\, \varphi$ (resp., $\psi = \langle E \rangle\, \varphi$) can be dealt with as $\psi = \langle A \rangle\, \varphi$ (resp., $\psi = \langle B \rangle\, \varphi$).

(Completeness) The proof is by induction on the structure of the formula $\psi$.

- $\psi$ is a Boolean combination of propositions $\beta$: if $\rho$ is associated with $d$ and $\mathcal{K}, \rho \models \beta$, then $VAL(\beta, d) = \top$ and thus check($\mathcal{K}, \psi, d$) has a successful computation.
- $\psi = \varphi_1 \vee \varphi_2$: if there exists a track $\rho$ associated with $d$ such that $\mathcal{K}, \rho \models \varphi_1 \vee \varphi_2$, then $\mathcal{K}, \rho \models \varphi_i$, for some $i \in \{1, 2\}$. By the inductive hypothesis, check($\mathcal{K}, \varphi_i, d$) has a successful computation, and hence check($\mathcal{K}, \psi, d$) has a successful computation.
- $\psi = \langle A \rangle\, \varphi$: if there exists a track $\rho$, associated with $d$, such that $\mathcal{K}, \rho \models \langle A \rangle\, \varphi$, then, by definition, there

**Algorithm 2** ProvideCounterex($\mathcal{K}, \psi$)

$(v_{in}, S, v_{fin}) \leftarrow$ aDescrEl($\mathcal{K}, w_0$, FORW)
**return** Check($\mathcal{K}$, to$\exists$A$\overline{\text{A}}$BE($\neg\psi$), $(v_{in}, S, v_{fin})$)

exists a track $\overline{\rho}$, with fst($\overline{\rho}$) = lst($\rho$) = $v_{fin}$, such that $\mathcal{K}, \overline{\rho} \models \varphi$. If $d' = (v_{fin}, S', v'_{fin})$ is the descriptor element for $\overline{\rho}$, then, by the inductive hypothesis, check($\mathcal{K}, \varphi, d'$) has a successful computation. Since there exists a computation where the non-deterministic call to aDescrEl($\mathcal{K}, v_{fin}$, FORW) returns the descriptor element $d'$ for $\overline{\rho}$, it follows that check($\mathcal{K}, \psi, d$) has a successful computation.

- $\psi = \langle \text{B} \rangle \varphi$: if there exists a track $\rho$, associated with $d$, such that $\mathcal{K}, \rho \models \langle \text{B} \rangle \varphi$, there are two possible cases.
  (i) $\mathcal{K}, \overline{\rho} \models \varphi$, with $\rho = \overline{\rho} \cdot v_{fin}$ for some $\overline{\rho} \in \text{Trk}_\mathcal{K}$. If $d' = (v_{in}, S', v'_{fin})$ is the descriptor element for $\overline{\rho}$, by the inductive hypothesis, check($\mathcal{K}, \varphi, d'$) has a successful computation. Since there is a computation where aDescrEl($\mathcal{K}, v_{in}$, FORW) returns $d'$ and both $(v'_{fin}, v_{fin}) \in \delta$ and $(v_{in}, S' \cup \{v'_{fin}\}, v_{fin}) = d$, it follows that check($\mathcal{K}, \psi, d$) has a successful computation.
  (ii) $\mathcal{K}, \overline{\rho} \models \varphi$ with $\rho = \overline{\rho} \cdot \tilde{\rho}$ for some $\overline{\rho}, \tilde{\rho} \in \text{Trk}_\mathcal{K}$. Let $d' = (v_{in}, S', v'_{fin})$ and $d'' = (v''_{in}, S'', v''_{fin})$ be the descriptor elements for $\overline{\rho}$ and $\tilde{\rho}$, respectively. Obviously, it holds that concat($d', d''$) = $d$. By the inductive hypothesis, check($\mathcal{K}, \varphi, d'$) has a successful computation. Since both $\overline{\rho}$ and $\tilde{\rho}$ are witnessed, there is a computation where the calls to aDescrEl($\mathcal{K}, v_{in}$, FORW) and aDescrEl($\mathcal{K}, v''_{in}$, FORW) non-deterministically return $d'$ and $d''$, respectively, and $(v'_{fin}, v''_{in}) \in \delta$ is non-deterministically chosen. Hence, check($\mathcal{K}, \psi, d$) has a successful computation.

The case $\psi = \langle \overline{\text{A}} \rangle \varphi$ (resp., $\psi = \langle \overline{\text{E}} \rangle \varphi$) can be dealt with as $\psi = \langle \text{A} \rangle \varphi$ (resp., $\psi = \langle \text{B} \rangle \varphi$). ∎

It is worth pointing out that Check($\mathcal{K}, \psi, d$) cannot deal with $\langle \overline{\text{B}} \rangle$ and $\langle \overline{\text{E}} \rangle$ modalities. In [14], to cope with them, Molinari et al. introduced the notion of *track descriptor*.

The procedure ProvideCounterex($\mathcal{K}, \psi$) (Algorithm 2) has a successful computation iff $\mathcal{K} \not\models \psi$, where $\psi$ is a $\forall$A$\overline{\text{A}}$BE formula, to$\exists$A$\overline{\text{A}}$BE($\neg\psi$) is the $\exists$A$\overline{\text{A}}$BE formula equivalent to $\neg\psi$, and $w_0$ is the initial state of $\mathcal{K}$.

On the one hand, if ProvideCounterex($\mathcal{K}, \psi$) has a successful computation, then there exists a witnessed descriptor element $d = (v_{in}, S, v_{fin})$, where $v_{in}$ is $w_0$ (the initial state of $\mathcal{K}$), such that check($\mathcal{K}$, to$\exists$A$\overline{\text{A}}$BE($\neg\psi$), $d$) has a successful computation. This means that there exists a track $\rho$, associated with $d$, such that $\mathcal{K}, \rho \models \neg\psi$, and thus $\mathcal{K} \not\models \psi$. On the other hand, if $\mathcal{K} \not\models \psi$, then there exists an initial track $\rho$ such that $\mathcal{K}, \rho \models \neg\psi$. Let $d$ be the descriptor element for $\rho$; check($\mathcal{K}$, to$\exists$A$\overline{\text{A}}$BE($\neg\psi$), $d$) has a successful computation: some non-deterministic instance of aDescrEl($\mathcal{K}, w_0$, FORW) returns $d$ (as it is witnessed

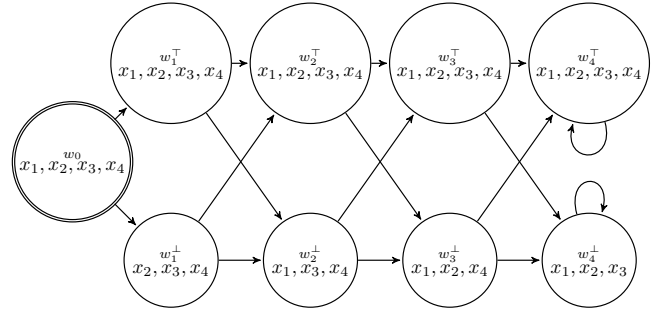

Figure 4. Kripke structure $\mathcal{K}_{SAT}^{Var}$ associated with a SAT formula with variables $Var = \{x_1, x_2, x_3, x_4\}$.

by an initial track), and thus ProvideCounterex($\mathcal{K}, \psi$) has a successful computation.

As for the complexity ProvideCounterex($\mathcal{K}, \psi$) runs in non-deterministic polynomial time (it is in NP) since the number of recursive invocations of the procedure Check is $O(|\psi|)$ and each invocation requires time polynomial in $|W|$ while generating descriptor elements. Therefore, the model checking problem for $\forall$A$\overline{\text{A}}$BE belongs to coNP.

We conclude the section by proving that the model checking problem for $\forall$A$\overline{\text{A}}$BE is coNP-complete. Such a result is an easy corollary of the following theorem.

*Theorem 2:* Let $\mathcal{K}$ be a Kripke structure and $\beta \in$ Prop be a Boolean combination of proposition letters. The problem of deciding whether $\mathcal{K} \not\models \beta$ is NP-hard (under a LOGSPACE reduction).

*Proof:* We provide a LOGSPACE reduction from the NP-complete SAT problem to the considered problem. Let $\beta$ be a Boolean formula over a set of variables $Var = \{x_1, \ldots, x_n\}$. We build a Kripke structure, $\mathcal{K}_{SAT}^{Var} = (\mathcal{AP}, W, \delta, \mu, w_0)$, with:

- $\mathcal{AP} = Var$;
- $W = \{w_0\} \cup \{w_i^\ell \mid \ell \in \{\top, \bot\}, 1 \leq i \leq n\}$;
- $\delta = \{(w_0, w_1^\top), (w_0, w_1^\bot)\} \cup \{(w_n^\top, w_n^\top), (w_n^\bot, w_n^\bot)\} \cup \{(w_i^\ell, w_{i+1}^m) \mid \ell, m \in \{\top, \bot\}, 1 \leq i \leq n-1\}$;
- $\mu(w_0) = \mathcal{AP}$;
- for $1 \leq i \leq n$, $\mu(w_i^\top) = \mathcal{AP}$ and $\mu(w_i^\bot) = \mathcal{AP} \setminus \{x_i\}$.

See Fig. 4 for an example of $\mathcal{K}_{SAT}^{Var}$, for $Var = \{x_1, .., x_4\}$.

It is immediate to see that any initial track $\rho$ of any length induces a truth assignment to the variables of $Var$: for any $x_i \in Var$, $x_i$ evaluates to $\top$ iff $x_i \in \bigcap_{w \in \text{states}(\rho)} \mu(w)$. Vice versa, for any possible truth assignment to the variables in $Var$, there exists an initial track $\rho$ that induces such an assignment: we include in the track the state $w_i^\top$ if $x_i$ is assigned to $\top$, $w_i^\bot$ otherwise.

Let $\gamma = \neg\beta$. It holds that $\beta$ is satisfiable iff there exists an initial track $\rho \in \text{Trk}_{\mathcal{K}_{SAT}^{Var}}$ such that $\mathcal{K}_{SAT}^{Var}, \rho \models \beta$, that is, iff $\mathcal{K}_{SAT}^{Var} \not\models \gamma$. To conclude, we observe that $\mathcal{K}_{SAT}^{Var}$ can be built with logarithmic working space. ∎

It immediately follows that checking whether $\mathcal{K} \not\models \beta$ for $\beta \in$ Prop is NP-complete, so model checking for formulas of Prop is coNP-complete. Moreover, since a Boolean combination of proposition letters in Prop is also a $\forall A\overline{A}BE$ formula, `ProvideCounterex`$(\mathcal{K},\psi)$ is at least as hard as checking whether $\mathcal{K} \not\models \beta$ for $\beta \in$ Prop. Thus, `ProvideCounterex`$(\mathcal{K},\psi)$ is NP-complete, hence the model checking problem for $\forall A\overline{A}BE$ is coNP-complete.

From the lower bound for Prop, it immediately follows that model checking for $A\overline{A}$ is coNP-hard (and we already know from [15] that it is in PSPACE).

## V. PSPACE-HARDNESS OF THE FRAGMENT $A\overline{A}B\overline{E}$

In [15], Molinari et al. showed how to extract from the proposed model checking algorithm for $A\overline{A}B\overline{B}\overline{E}$ a model checking algorithm for $A\overline{A}B\overline{E}$ which works in polynomial (not exponential) space. It benefits from the fact that $A\overline{A}B\overline{E}$ lacks the $B$ modality. Here, we prove that there is no way to improve such an algorithm by showing that model checking for $A\overline{B}$ is a PSPACE-hard problem (Theorem 3). PSPACE-completeness of $A\overline{A}B\overline{E}$ (and $A\overline{B}$) immediately follows. As a by-product, model checking for $A\overline{A}B\overline{B}\overline{E}$ is PSPACE-hard as well (in [15], we only proved that it is NP-hard in the non-succinct case).

We provide a reduction from the QBF problem (i.e., the problem of determining the truth of a *fully-quantified* Boolean formula in *prenex normal form*)—which is known to be PSPACE-complete (see, for example, [22])—to the model checking problem for $A\overline{B}$ formulas over finite Kripke structures. We consider a quantified Boolean formula $\psi = Q_n x_n Q_{n-1} x_{n-1} \cdots Q_1 x_1 \phi(x_n, x_{n-1}, \cdots, x_1)$, where $Q_i \in \{\exists, \forall\}$, for $i = 1, \cdots, n$, and $\phi(x_n, x_{n-1}, \cdots, x_1)$ is a quantifier-free Boolean formula. Let $Var = \{x_n, \ldots, x_1\}$ be the set of variables of $\psi$. We define the Kripke structure $\mathcal{K}_{QBF}^{Var} = (\mathcal{AP}, W, \delta, \mu, w_0)$ as follows:

- $\mathcal{AP} = Var \cup \{start\} \cup \{x_{i\,aux} \mid 1 \leq i \leq n\}$;
- $W = \{w_{x_i}^\ell \mid 1 \leq i \leq n,\ \ell \in \{\bot_1, \bot_2, \top_1, \top_2\}\} \cup \{w_0, w_1, sink\}$;
- if $n = 0$, $\delta = \{(w_0, w_1), (w_1, sink), (sink, sink)\}$;
  if $n > 0$,
  $\delta = \{(w_0, w_1), (w_1, w_{x_n}^{\top_1}), (w_1, w_{x_n}^{\bot_1})\} \cup$
  $\{(w_{x_i}^{\top_1}, w_{x_i}^{\top_2}), (w_{x_i}^{\bot_1}, w_{x_i}^{\bot_2}) \mid 1 \leq i \leq n\} \cup$
  $\{(w_{x_i}^\ell, w_{x_{i-1}}^m) \mid \ell \in \{\bot_2, \top_2\}, m \in \{\bot_1, \top_1\}, n \leq i \leq 2\} \cup$
  $\{(w_{x_1}^{\top_2}, sink), (w_{x_1}^{\bot_2}, sink), (sink, sink)\}$.
- $\mu(w_0) = \mu(w_1) = Var \cup \{start\}$;
  $\mu(w_{x_i}^\ell) = Var \cup \{x_{i\,aux}\}$, for $1 \leq i \leq n$, $\ell \in \{\top_1, \top_2\}$;
  $\mu(w_{x_i}^\ell) = (Var \setminus \{x_i\}) \cup \{x_{i\,aux}\}$, for $1 \leq i \leq n$ and $\ell \in \{\bot_1, \bot_2\}$;
  $\mu(sink) = Var$.

An example of such a Kripke structure where $Var = \{x, y, z\}$ is given in Figure 5.

From $\psi$, we obtain the $A\overline{B}$ formula $\xi = start \rightarrow \xi_n$, where

$$\xi_i = \begin{cases} \phi(x_n, x_{n-1}, \cdots x_1) & i = 0 \\ \langle \overline{B} \rangle \big( (\langle A \rangle x_{i\,aux}) \wedge \xi_{i-1} \big) & i > 0 \wedge Q_i = \exists \\ [\overline{B}] \big( (\langle A \rangle x_{i\,aux}) \rightarrow \xi_{i-1} \big) & i > 0 \wedge Q_i = \forall \end{cases}.$$

Both $\mathcal{K}_{QBF}^{Var}$ and $\xi$ can be built by using logarithmic working space. We will show (proof of Theorem 3) that $\psi$ is true if and only if $\mathcal{K}_{QBF}^{Var} \models \xi$.

As a preliminary step, we introduce some technical definitions and prove the auxiliary Lemma 1.

Given a Kripke structure $\mathcal{K} = (\mathcal{AP}, W, \delta, \mu, w_0)$ and an $A\overline{B}$ formula $\psi$, we denote by $p\ell(\psi)$ the set of proposition letters occurring in $\psi$ and by $\mathcal{K}_{|p\ell(\psi)}$ the structure obtained from $\mathcal{K}$ by restricting the labelling of each state to $p\ell(\psi)$, namely, the Kripke structure $(\overline{\mathcal{AP}}, W, \delta, \overline{\mu}, w_0)$, where $\overline{\mathcal{AP}} = \mathcal{AP} \cap p\ell(\psi)$ and $\overline{\mu}(w) = \mu(w) \cap p\ell(\psi)$, for all $w \in W$. Moreover, for $v \in W$, we denote by $reach(\mathcal{K}, v)$ the subgraph of $\mathcal{K}$, with $v$ as its initial state, consisting of all and only the states which are reachable from $v$, namely, the Kripke structure $(\mathcal{AP}, W', \delta', \mu', v)$, where $W' = \{w \in W \mid$ there exists $\rho \in \text{Trk}_\mathcal{K}$ with $\text{fst}(\rho) = v$ and $\text{lst}(\rho) = w\}$, $\delta' = \delta \cap (W' \times W')$, and $\mu'(w) = \mu(w)$, for all $w \in W'$. As usual, we say that two Kripke structures $\mathcal{K} = (\mathcal{AP}, W, \delta, \mu, w_0)$ and $\mathcal{K}' = (\mathcal{AP}', W', \delta', \mu', w_0')$ are *isomorphic* ($\mathcal{K} \sim \mathcal{K}'$ for short) iff there is a *bijection* $f : W \mapsto W'$ such that (i) $f(w_0) = w_0'$; (ii) for all $u, v \in W$, $(u, v) \in \delta$ iff $(f(u), f(v)) \in \delta'$; (iii) for all $v \in W$, $\mu(v) = \mu'(f(v))$. Finally, if $\mathcal{A}_\mathcal{K} = (\mathcal{AP}, \mathbb{I}, A_\mathbb{I}, B_\mathbb{I}, E_\mathbb{I}, \sigma)$ is the abstract interval model induced by a Kripke structure $\mathcal{K}$ and $\rho \in \text{Trk}_\mathcal{K}$, we denote $\sigma(\rho)$ by $\mathcal{L}(\mathcal{K}, \rho)$.

Let $\mathcal{K}$ and $\mathcal{K}'$ be two Kripke structures. The following lemma states that, for any $A\overline{B}$ formula $\psi$, if the same set of proposition letters, restricted to $p\ell(\psi)$, holds over two tracks $\rho \in \text{Trk}_\mathcal{K}$ and $\rho' \in \text{Trk}_{\mathcal{K}'}$, and the subgraphs consisting of the states reachable from, respectively, $\text{lst}(\rho)$ and $\text{lst}(\rho')$ are isomorphic, then $\rho$ and $\rho'$ are equivalent with respect to $\psi$.

*Lemma 1:* Given an $A\overline{B}$ formula $\psi$, two Kripke structures $\mathcal{K} = (\mathcal{AP}, W, \delta, \mu, w_0)$ and $\mathcal{K}' = (\mathcal{AP}', W', \delta', \mu', w_0')$, and two tracks $\rho \in \text{Trk}_\mathcal{K}$ and $\rho' \in \text{Trk}_{\mathcal{K}'}$ such that

$$\mathcal{L}(\mathcal{K}_{|p\ell(\psi)}, \rho) = \mathcal{L}(\mathcal{K}'_{|p\ell(\psi)}, \rho') \quad \text{and}$$

$$reach(\mathcal{K}_{|p\ell(\psi)}, \text{lst}(\rho)) \sim reach(\mathcal{K}'_{|p\ell(\psi)}, \text{lst}(\rho')),$$

it holds that $\mathcal{K}, \rho \models \psi \iff \mathcal{K}', \rho' \models \psi$.

*Proof:* The proof is by structural induction on $\psi$.

- $\psi = p$, with $p \in \mathcal{AP}$ ($p\ell(p) = \{p\}$). If $\mathcal{K}, \rho \models p$, then $p \in \mathcal{L}(\mathcal{K}, \rho)$ and hence $p \in \mathcal{L}(\mathcal{K}_{|p\ell(\psi)}, \rho)$. It immediately follows that $p \in \mathcal{L}(\mathcal{K}'_{|p\ell(\psi)}, \rho')$, and thus $p \in \mathcal{L}(\mathcal{K}', \rho')$ and $\mathcal{K}', \rho' \models p$.
- $\psi = \neg \phi$ ($p\ell(\phi) = p\ell(\psi)$). If $\mathcal{K}, \rho \models \neg \phi$, then $\mathcal{K}, \rho \not\models \phi$. By the inductive hypothesis, $\mathcal{K}', \rho' \not\models \phi$ and thus $\mathcal{K}', \rho' \models \neg \phi$.
- $\psi = \phi_1 \wedge \phi_2$. If $\mathcal{K}, \rho \models \phi_1 \wedge \phi_2$, then $\mathcal{K}, \rho \models \phi_1$. Since, by hypothesis, $\mathcal{L}(\mathcal{K}_{|p\ell(\psi)}, \rho) = \mathcal{L}(\mathcal{K}'_{|p\ell(\psi)}, \rho')$ and $reach(\mathcal{K}_{|p\ell(\psi)}, \operatorname{lst}(\rho)) \sim reach(\mathcal{K}'_{|p\ell(\psi)}, \operatorname{lst}(\rho'))$, it holds that $\mathcal{L}(\mathcal{K}_{|p\ell(\phi_1)}, \rho) = \mathcal{L}(\mathcal{K}'_{|p\ell(\phi_1)}, \rho')$ and $reach(\mathcal{K}_{|p\ell(\phi_1)}, \operatorname{lst}(\rho)) \sim reach(\mathcal{K}'_{|p\ell(\phi_1)}, \operatorname{lst}(\rho'))$, since $p\ell(\phi_1) \subseteq p\ell(\psi)$. By the inductive hypothesis, $\mathcal{K}', \rho' \models \phi_1$. The same argument works for $\phi_2$. The thesis follows.

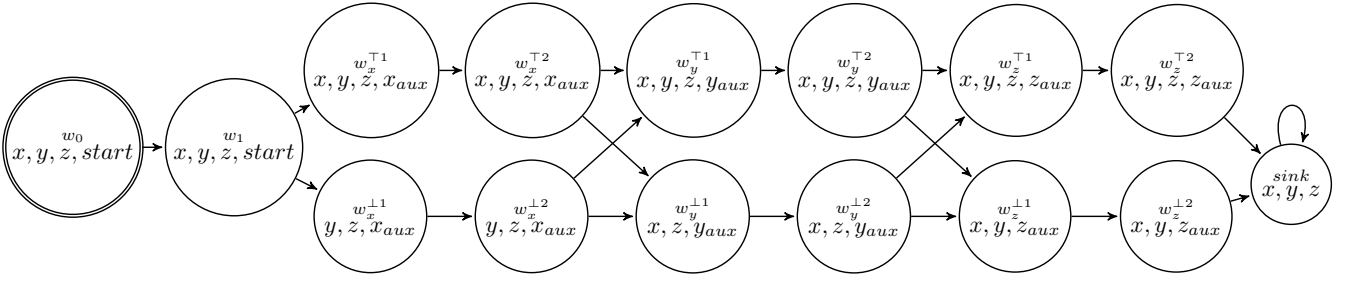

Figure 5. Kripke structure $\mathcal{K}_{QBF}^{x,y,z}$ associated with a quantified Boolean formula with variables $x$, $y$, $z$.

- $\psi = \langle A \rangle \phi$. If it holds that $\mathcal{K}, \rho \models \langle A \rangle \phi$, then there exists a track $\overline{\rho} \in \operatorname{Trk}_{\mathcal{K}}$ such that $\operatorname{fst}(\overline{\rho}) = \operatorname{lst}(\rho)$ and $\mathcal{K}, \overline{\rho} \models \phi$, with $p\ell(\phi) = p\ell(\psi)$. By hypothesis, $reach(\mathcal{K}_{|p\ell(\psi)}, \operatorname{lst}(\rho)) \sim reach(\mathcal{K}'_{|p\ell(\psi)}, \operatorname{lst}(\rho'))$. Hence, there exists a track $\overline{\rho}' \in \operatorname{Trk}_{\mathcal{K}'}$, with $\operatorname{fst}(\overline{\rho}') = \operatorname{lst}(\rho')$, such that $|\overline{\rho}| = |\overline{\rho}'|$ and for all $0 \leq i \leq |\overline{\rho}| - 1$, $f(\overline{\rho}(i)) = \overline{\rho}'(i)$, where $f$ is the (an) isomorphism between $reach(\mathcal{K}_{|p\ell(\psi)}, \operatorname{lst}(\rho))$ and $reach(\mathcal{K}'_{|p\ell(\psi)}, \operatorname{lst}(\rho'))$. It immediately follows that $\mathcal{L}(\mathcal{K}_{|p\ell(\phi)}, \overline{\rho}) = \mathcal{L}(\mathcal{K}'_{|p\ell(\phi)}, \overline{\rho}')$.

We now prove that $reach(\mathcal{K}_{|p\ell(\phi)}, \operatorname{lst}(\overline{\rho}))$ is isomorphic to $reach(\mathcal{K}'_{|p\ell(\phi)}, \operatorname{lst}(\overline{\rho}'))$. To this end, it suffices to prove that the restriction of the isomorphism $f$ to the states of $reach(\mathcal{K}_{|p\ell(\phi)}, \operatorname{lst}(\overline{\rho}))$, say $f'$, is an isomorphism between $reach(\mathcal{K}_{|p\ell(\phi)}, \operatorname{lst}(\overline{\rho}))$ and $reach(\mathcal{K}'_{|p\ell(\phi)}, \operatorname{lst}(\overline{\rho}'))$ (notice that the Kripke structure $reach(\mathcal{K}_{|p\ell(\phi)}, \operatorname{lst}(\overline{\rho}))$ is a subgraph of $reach(\mathcal{K}_{|p\ell(\psi)}, \operatorname{lst}(\rho))$). First, it holds that $f(\operatorname{lst}(\overline{\rho})) = f'(\operatorname{lst}(\overline{\rho})) = \operatorname{lst}(\overline{\rho}')$. Next, if $w$ is any state of $reach(\mathcal{K}_{|p\ell(\phi)}, \operatorname{lst}(\overline{\rho}))$, then $f(w) = f'(w) = w'$ is a state of $reach(\mathcal{K}'_{|p\ell(\phi)}, \operatorname{lst}(\overline{\rho}'))$, as from the existence of a track from $\operatorname{lst}(\overline{\rho})$ to $w$, it follows that there is an isomorphic track (w.r.t. $f$) from $\operatorname{lst}(\overline{\rho}')$ to $w'$. Moreover, if $(w, \overline{w}) \in \delta$, then $\overline{w}$ belongs to $reach(\mathcal{K}_{|p\ell(\phi)}, \operatorname{lst}(\overline{\rho}))$, and thus $(w', f(\overline{w})) \in \delta'$ and $f(\overline{w}) = f'(\overline{w})$ belongs to $reach(\mathcal{K}'_{|p\ell(\phi)}, \operatorname{lst}(\overline{\rho}'))$. We can thus conclude that, for any two states $v, v'$ of $reach(\mathcal{K}_{|p\ell(\phi)}, \operatorname{lst}(\overline{\rho}))$, it holds that $(v, v')$ is an edge iff $(f'(v), f'(v'))$ is an edge of $reach(\mathcal{K}'_{|p\ell(\phi)}, \operatorname{lst}(\overline{\rho}'))$. By the inductive hypothesis, $\mathcal{K}', \overline{\rho}' \models \phi$ and hence $\mathcal{K}', \rho' \models \langle A \rangle \phi$.

- $\psi = \langle \overline{B} \rangle \phi$. If $\mathcal{K}, \rho \models \langle \overline{B} \rangle \phi$, then $\mathcal{K}, \rho \cdot \overline{\rho} \models \phi$, with $p\ell(\psi) = p\ell(\phi)$, where $\rho \cdot \overline{\rho} \in \operatorname{Trk}_{\mathcal{K}}$ and $\overline{\rho}$ is either a single state or a proper track. In analogy to the previous case, let $\overline{\rho}' \in \operatorname{Trk}_{\mathcal{K}'}$ such that $|\overline{\rho}| = |\overline{\rho}'|$ and, for all $0 \leq i < |\overline{\rho}|$, $f(\overline{\rho}(i)) = \overline{\rho}'(i)$, where $f$ is the isomorphism between $reach(\mathcal{K}_{|p\ell(\psi)}, \operatorname{lst}(\rho))$ and $reach(\mathcal{K}'_{|p\ell(\psi)}, \operatorname{lst}(\rho'))$. Since $f(\operatorname{lst}(\rho)) = \operatorname{lst}(\rho')$, by definition of isomorphism, $(\operatorname{lst}(\rho), \operatorname{fst}(\overline{\rho})) \in \delta$ implies $(\operatorname{lst}(\rho'), \operatorname{fst}(\overline{\rho}')) \in \delta'$. It follows that $\mathcal{L}(\mathcal{K}_{|p\ell(\phi)}, \overline{\rho}) = \mathcal{L}(\mathcal{K}'_{|p\ell(\phi)}, \overline{\rho}')$ and $reach(\mathcal{K}_{|p\ell(\phi)}, \operatorname{lst}(\overline{\rho}))$ is isomorphic to $reach(\mathcal{K}'_{|p\ell(\phi)}, \operatorname{lst}(\overline{\rho}'))$. Finally,

$$\mathcal{L}(\mathcal{K}_{|p\ell(\phi)}, \rho \cdot \overline{\rho}) = \mathcal{L}(\mathcal{K}_{|p\ell(\phi)}, \rho) \cap \mathcal{L}(\mathcal{K}_{|p\ell(\phi)}, \overline{\rho}) = \mathcal{L}(\mathcal{K}'_{|p\ell(\phi)}, \rho') \cap \mathcal{L}(\mathcal{K}'_{|p\ell(\phi)}, \overline{\rho}') = \mathcal{L}(\mathcal{K}'_{|p\ell(\phi)}, \rho' \cdot \overline{\rho}')$$

and $reach(\mathcal{K}_{|p\ell(\phi)}, \operatorname{lst}(\rho \cdot \overline{\rho}))$ is isomorphic to $reach(\mathcal{K}'_{|p\ell(\phi)}, \operatorname{lst}(\rho' \cdot \overline{\rho}'))$. By the inductive hypothesis, $\mathcal{K}', \rho' \cdot \overline{\rho}' \models \phi$ and therefore $\mathcal{K}', \rho' \models \langle \overline{B} \rangle \phi$. ∎

*Theorem 3:* The model checking problem for A$\overline{B}$ over finite Kripke structures is PSPACE-hard (under LOGSPACE reductions).

*Proof:* We prove that the quantified Boolean formula $\psi = Q_n x_n Q_{n-1} x_{n-1} \cdots Q_1 x_1 \phi(x_n, x_{n-1}, \cdots x_1)$ is true iff $\mathcal{K}_{QBF}^{x_n, \cdots, x_1} \models \xi$ by induction on the number of variables $n \in \mathbb{N}$ of $\psi$. In the following, $\phi(x_n, x_{n-1}, \cdots x_1)\{x_i/v\}$, with $v \in \{\top, \bot\}$, denotes the formula obtained from $\phi(x_n, x_{n-1}, \cdots x_1)$ by replacing all occurrences of $x_i$ by $v$. It is worth noticing that $\mathcal{K}_{QBF}^{x_n, x_{n-1}, \cdots, x_1}$ and $\mathcal{K}_{QBF}^{x_{n-1}, \cdots, x_1}$ are isomorphic when they are restricted to the states $w_{x_{n-1}}^{\top 1}$, $w_{x_{n-1}}^{\top 2}, w_{x_{n-1}}^{\bot 1}, w_{x_{n-1}}^{\bot 2}, \cdots, w_{x_1}^{\top 1}, w_{x_1}^{\top 2}, w_{x_1}^{\bot 1}, w_{x_1}^{\bot 2}, sink$ (i.e., the leftmost part of both Kripke structures is eliminated), and the labelling of states is suitably restricted as well. Moreover, only the track $w_0 w_1$ satisfies $start$ and, for $i = n, \cdots, 1$, the proposition letter $x_{i\,aux}$ is satisfied by the two tracks $w_{x_i}^{\top 1} w_{x_i}^{\top 2}$ and $w_{x_i}^{\bot 1} w_{x_i}^{\bot 2}$ only.

(Case $n = 0$) $\psi$ equals $\phi$ and it has no variables. The states of $\mathcal{K}_{QBF}^{\emptyset}$ are $W = \{w_0, w_1, sink\}$ and $\xi = start \to \phi$.

Let us assume $\phi$ to be true. All initial tracks of length greater than 2 trivially satisfy $\xi$, as $start$ does not hold on them. As for $w_0 w_1$, it is true that $\mathcal{K}_{QBF}^{\emptyset}, w_0 w_1 \models \phi$, since $\phi$ is true (its truth does not depend on the proposition letters that hold on $w_0 w_1$, because it has no variables). Thus $\mathcal{K}_{QBF}^{\emptyset} \models \xi$. Vice versa, if $\mathcal{K}_{QBF}^{\emptyset} \models \xi$, then, in particular, $\mathcal{K}_{QBF}^{\emptyset}, w_0 w_1 \models \phi$. But $\phi$ has no variables, hence it is true.

(Case $n \geq 1$) Let us consider the quantified Boolean formula $\psi = Q_n x_n Q_{n-1} x_{n-1} \cdots Q_1 x_1 \phi(x_n, x_{n-1}, \cdots x_1)$. We distinguish two cases, depending on whether $Q_n = \exists$ or $Q_n = \forall$, and for both we prove the two implications.

○ Case $Q_n = \exists$:

($\Rightarrow$) If the formula $\psi$ is true, then, by definition, there exists $\upsilon \in \{\top, \bot\}$ such that if we replace all occurrences of $x_n$ in $\phi(x_n, x_{n-1}, \cdots x_1)$ by $\upsilon$, we get the formula $\phi'(x_{n-1}, \cdots x_1) = \phi(x_n, x_{n-1}, \cdots x_1)\{x_n/\upsilon\}$ such that $\psi' = Q_{n-1} x_{n-1} \cdots Q_1 x_1 \phi'(x_{n-1}, \cdots x_1)$ is a true quantified Boolean formula. By the inductive hypothesis $\mathcal{K}_{QBF}^{x_{n-1},\cdots,x_1} \models \xi'$, where $\xi' = start \to \xi'_{n-1}$ is obtained from $\psi'$ and $\xi'_{n-1} = \xi_{n-1}\{x_n/\upsilon\}$. It follows that $\mathcal{K}_{QBF}^{x_{n-1},\cdots,x_1}, w'_0 w'_1 \models \xi'_{n-1}$, where $w'_0$ and $w'_1$ are the two "leftmost" states of $\mathcal{K}_{QBF}^{x_{n-1},\cdots,x_1}$ (corresponding to $w_0$ and $w_1$ of $\mathcal{K}_{QBF}^{x_n,\cdots,x_1}$).

We prove that $\mathcal{K}_{QBF}^{x_n,\cdots,x_1} \models \xi$. Let us consider a generic initial track $\rho$ in $\mathcal{K}_{QBF}^{x_n,\cdots,x_1}$. If it does not satisfy $start$, then it trivially holds that $\mathcal{K}_{QBF}^{x_n,\cdots,x_1}, \rho \models \xi$. Otherwise ($\rho = w_0 w_1$), we have to show that $\mathcal{K}_{QBF}^{x_n,\cdots,x_1}, w_0 w_1 \models \langle \overline{B} \rangle ((\langle A \rangle x_{n\,aux}) \wedge \xi_{n-1}) \ (= \xi_n)$. If $\upsilon = \top$, we consider $w_0 w_1 w_{x_n}^{\top 1}$; otherwise, we consider $w_0 w_1 w_{x_n}^{\bot 1}$. In the first case (the other is symmetric), we must prove that $\mathcal{K}_{QBF}^{x_n,\cdots,x_1}, w_0 w_1 w_{x_n}^{\top 1} \models (\langle A \rangle x_{n\,aux}) \wedge \xi_{n-1}$. It trivially holds that $\mathcal{K}_{QBF}^{x_n,\cdots,x_1}, w_0 w_1 w_{x_n}^{\top 1} \models \langle A \rangle x_{n\,aux}$. Hence, we only need to show that $\mathcal{K}_{QBF}^{x_n,\cdots,x_1}, w_0 w_1 w_{x_n}^{\top 1} \models \xi_{n-1}$.

As we have shown, by the inductive hypothesis, it holds that $\mathcal{K}_{QBF}^{x_{n-1},\cdots,x_1}, w'_0 w'_1 \models \xi'_{n-1}(= \xi_{n-1}\{x_n/\top\})$. Now, since

- $p\ell(\xi_{n-1}\{x_n/\top\}) = \{x_1, \cdots, x_{n-1}, x_{1\,aux}, \cdots, x_{n-1\,aux}\}$,

- $\mathcal{L}(\mathcal{K}_{QBF}^{x_{n-1},\cdots,x_1}{}_{|p\ell(\xi_{n-1}\{x_n/\top\})}, w'_0 w'_1) = \{x_{n-1}, \cdots, x_1\}$,

- $\mathcal{L}(\mathcal{K}_{QBF}^{x_n,\cdots,x_1}{}_{|p\ell(\xi_{n-1}\{x_n/\top\})}, w_0 w_1 w_{x_n}^{\top 1} w_{x_n}^{\top 2}) = \{x_{n-1}, .., x_1\}$,

- $reach(\mathcal{K}_{QBF}^{x_n,\cdots,x_1}{}_{|p\ell(\xi_{n-1}\{x_n/\top\})}, w_{x_n}^{\top 2})$ is isomorphic to $reach(\mathcal{K}_{QBF}^{x_{n-1},\cdots,x_1}{}_{|p\ell(\xi_{n-1}\{x_n/\top\})}, w'_1)$,

by Lemma 1, $\mathcal{K}_{QBF}^{x_n,\cdots,x_1}, w_0 w_1 w_{x_n}^{\top 1} w_{x_n}^{\top 2} \models \xi'_{n-1}$. Hence, $\mathcal{K}_{QBF}^{x_n,\cdots,x_1}, w_0 w_1 w_{x_n}^{\top 1} w_{x_n}^{\top 2} \models \xi_{n-1}$ as $x_n$ is in the labelling of the track $w_0 w_1 w_{x_n}^{\top 1} w_{x_n}^{\top 2}$ and of any $\overline{\rho}$ such that $w_0 w_1 w_{x_n}^{\top 1} w_{x_n}^{\top 2} \in \text{Pref}(\overline{\rho})$.

Now, if $n = 1$, then $\xi_{n-1} = \phi(x_n)$ and it holds that $\mathcal{K}_{QBF}^{x_n,\cdots,x_1}, w_0 w_1 w_{x_n}^{\top 1} \models \xi_{n-1}$.

If $n > 1$, either $\xi_{n-1} = \langle \overline{B} \rangle ((\langle A \rangle x_{n-1\,aux}) \wedge \xi_{n-2})$ or $\xi_{n-1} = [\overline{B}]((\langle A \rangle x_{n-1\,aux}) \to \xi_{n-2})$.

In the first case, since $\mathcal{K}_{QBF}^{x_n,\cdots,x_1}, w_0 w_1 w_{x_n}^{\top 1} w_{x_n}^{\top 2} \models \langle \overline{B} \rangle ((\langle A \rangle x_{n-1\,aux}) \wedge \xi_{n-2})$, there are only two possibilities: $\mathcal{K}_{QBF}^{x_n,\cdots,x_1}, w_0 w_1 w_{x_n}^{\top 1} w_{x_n}^{\top 2} w_{x_{n-1}}^{\top 1} \models (\langle A \rangle x_{n-1\,aux}) \wedge \xi_{n-2}$ or $\mathcal{K}_{QBF}^{x_n,\cdots,x_1}, w_0 w_1 w_{x_n}^{\top 1} w_{x_n}^{\top 2} w_{x_{n-1}}^{\bot 1} \models (\langle A \rangle x_{n-1\,aux}) \wedge \xi_{n-2}$. In both cases, $\mathcal{K}_{QBF}^{x_n,\cdots,x_1}, w_0 w_1 w_{x_n}^{\top 1} \models \langle \overline{B} \rangle ((\langle A \rangle x_{n-1\,aux}) \wedge \xi_{n-2})$.

Else $\mathcal{K}_{QBF}^{x_n,\cdots,x_1}, w_0 w_1 w_{x_n}^{\top 1} w_{x_n}^{\top 2} \models [\overline{B}]((\langle A \rangle x_{n-1\,aux}) \to \xi_{n-2})$. It follows that $\mathcal{K}_{QBF}^{x_n,\cdots,x_1}, w_0 w_1 w_{x_n}^{\top 1} w_{x_n}^{\top 2} w_{x_{n-1}}^{\top 1} \models \xi_{n-2}$ and $\mathcal{K}_{QBF}^{x_n,\cdots,x_1}, w_0 w_1 w_{x_n}^{\top 1} w_{x_n}^{\top 2} w_{x_{n-1}}^{\bot 1} \models \xi_{n-2}$. As a consequence, $\mathcal{K}_{QBF}^{x_n,\cdots,x_1}, w_0 w_1 w_{x_n}^{\top 1} \models [\overline{B}]((\neg \langle A \rangle x_{n-1\,aux}) \vee \xi_{n-2})$ ($= \xi_{n-1}$) (recall that the only successor of $w_{x_n}^{\top 1}$ in $\mathcal{K}_{QBF}^{x_n,\cdots,x_1}$ is $w_{x_n}^{\top 2}$ and, in particular, $\mathcal{K}_{QBF}^{x_n,\cdots,x_1}, w_0 w_1 w_{x_n}^{\top 1} w_{x_n}^{\top 2} \models \neg \langle A \rangle x_{n-1\,aux}$).

($\Leftarrow$) If we have $\mathcal{K}_{QBF}^{x_n,\cdots,x_1} \models \xi$, it holds that $\mathcal{K}_{QBF}^{x_n,\cdots,x_1}, w_0 w_1 \models \langle \overline{B} \rangle ((\langle A \rangle x_{n\,aux}) \wedge \xi_{n-1})$. Hence, either $\mathcal{K}_{QBF}^{x_n,\cdots,x_1}, w_0 w_1 w_{x_n}^{\top 1} \models (\langle A \rangle x_{n\,aux}) \wedge \xi_{n-1}$ or $\mathcal{K}_{QBF}^{x_n,\cdots,x_1}, w_0 w_1 w_{x_n}^{\bot 1} \models (\langle A \rangle x_{n\,aux}) \wedge \xi_{n-1}$. Let us consider the first case (the other is symmetric). It holds that $\mathcal{K}_{QBF}^{x_n,\cdots,x_1}, w_0 w_1 w_{x_n}^{\top 1} \models \xi_{n-1}\{x_n/\top\}$ and $\mathcal{K}_{QBF}^{x_n,\cdots,x_1}, w_0 w_1 w_{x_n}^{\top 1} w_{x_n}^{\top 2} \models \xi_{n-1}\{x_n/\top\}$ (as before). By Lemma 1, $\mathcal{K}_{QBF}^{x_{n-1},\cdots,x_1}, w'_0 w'_1 \models \xi_{n-1}\{x_n/\top\}$ ($= \xi'_{n-1}$) and thus $\mathcal{K}_{QBF}^{x_{n-1},\cdots,x_1} \models start \to \xi'_{n-1}$, namely, $\mathcal{K}_{QBF}^{x_{n-1},\cdots,x_1} \models \xi'$. By the inductive hypothesis, $\psi' = Q_{n-1} x_{n-1} \cdots Q_1 x_1 \phi(x_n, x_{n-1}, \cdots, x_1)\{x_n/\top\}$ is true. Hence, $\psi = \exists x_n Q_{n-1} x_{n-1} \cdots Q_1 x_1 \phi(x_n, x_{n-1}, \cdots x_1)$ is true.

○ Case $Q_n = \forall$:

($\Rightarrow$) Assume that both $\psi' = Q_{n-1} x_{n-1} \cdots Q_1 x_1 \phi(x_n, x_{n-1}, \cdots, x_1)\{x_n/\top\}$ and $\psi'' = Q_{n-1} x_{n-1} \cdots Q_1 x_1 \phi(x_n, x_{n-1}, \cdots, x_1)\{x_n/\bot\}$ are true quantified Boolean formulas. We show that $\mathcal{K}_{QBF}^{x_n,\cdots,x_1}, w_0 w_1 \models [\overline{B}]((\langle A \rangle x_{n\,aux}) \to \xi_{n-1})$. To this end, it suffices to prove that both $\mathcal{K}_{QBF}^{x_n,\cdots,x_1}, w_0 w_1 w_{x_n}^{\top 1} \models \xi_{n-1}$ and $\mathcal{K}_{QBF}^{x_n,\cdots,x_1}, w_0 w_1 w_{x_n}^{\bot 1} \models \xi_{n-1}$. This can be shown exactly as in the $\exists$ case.

($\Leftarrow$) If $\mathcal{K}_{QBF}^{x_n,\cdots,x_1} \models \xi$, then $\mathcal{K}_{QBF}^{x_n,\cdots,x_1}, w_0 w_1 \models [\overline{B}]((\langle A \rangle x_{n\,aux}) \to \xi_{n-1})$. Hence, $\mathcal{K}_{QBF}^{x_n,\cdots,x_1}, w_0 w_1 w_{x_n}^{\top 1} \models \xi_{n-1}$ and $\mathcal{K}_{QBF}^{x_n,\cdots,x_1}, w_0 w_1 w_{x_n}^{\bot 1} \models \xi_{n-1}$. Reasoning as in the $\exists$ case and by applying the inductive hypothesis twice, we get that $Q_{n-1} x_{n-1} \cdots Q_1 x_1 \phi(x_n, x_{n-1}, \cdots, x_1)\{x_n/\top\}$ and $Q_{n-1} x_{n-1} \cdots Q_1 x_1 \phi(x_n, x_{n-1}, \cdots, x_1)\{x_n/\bot\}$ are true; thus $\forall x_n Q_{n-1} x_{n-1} \cdots Q_1 x_1 \phi(x_n, x_{n-1}, \cdots, x_1)$ is true. ■

## VI. CONCLUSIONS

In this paper, we identified some HS fragments, namely, $\forall A\overline{A}BE$, $A\overline{A}B\overline{E}$, and $A\overline{A}$, whose model checking problem turns out to be (computationally) much simpler than that of full HS and of other, already-studied fragments of it, and comparable to that of point-based temporal logics (as an example, the model checking problem for $A\overline{A}B\overline{E}$ has the same complexity as that for LTL). We also showed that these fragments are expressive enough to capture meaningful properties of state-transition systems, such as, for instance, mutual exclusion, state reachability, and non-starvation.

As for future work, we are currently exploring two main research directions. On the one hand, we are looking for other well-behaved fragments of HS; on the other hand, we are thinking of relaxing the homogeneity assumption to be able to deal with properties involving temporal aggregation and the like (e.g., a constraint on the average speed of a moving device during a given time period, that can be predicated of time intervals as a whole only). Work on Duration Calculus (DC) model checking seems to be relevant to this latter research direction. DC is an interval temporal logic endowed with the additional notion of state. Each state is denoted by a state expression and characterized by a duration. Recent results on DC model checking as well as an account of related work can be found in [9].

ACKNOWLEDGEMENTS

The work by Adriano Peron has been supported by the SHERPA collaborative project, which has received funding from the European Community 7th Framework Programme (FP7/2007-2013) under grant agreements ICT-600958. He is solely responsible for its content. The paper does not represent the opinion of the European Community and the Community is not responsible for any use that might be made of the information contained therein. The work by Angelo Montanari has been supported by the GNCS project *Algorithms to model check and synthesize safety-critical systems*. We would like to thank the reviewers for their useful comments.

REFERENCES

[1] J. F. Allen, "Maintaining knowledge about temporal intervals," *Communications of the ACM*, vol. 26(11), pp. 832–843, 1983.

[2] H. Bowman and S. J. Thompson, "A decision procedure and complete axiomatization of finite interval temporal logic with projection," *Journal of Logic and Computation*, vol. 13(2), pp. 195–239, 2003.

[3] D. Bresolin, D. Della Monica, V. Goranko, A. Montanari, and G. Sciavicco, "The dark side of interval temporal logic: marking the undecidability border," *Annals of Mathematics and Artificial Intelligence*, vol. 71(1-3), pp. 41–83, 2014.

[4] D. Bresolin, V. Goranko, A. Montanari, and P. Sala, "Tableau-based decision procedures for the logics of subinterval structures over dense orderings," *Journal of Logic and Computation*, vol. 20(1), pp. 133–166, 2010.

[5] D. Bresolin, V. Goranko, A. Montanari, and G. Sciavicco, "Propositional interval neighborhood logics: Expressiveness, decidability, and undecidable extensions," *Annals of Pure and Applied Logic*, vol. 161(3), pp. 289–304, 2009.

[6] D. Bresolin, A. Montanari, P. Sala, and G. Sciavicco, "What's decidable about Halpern and Shoham's interval logic? The maximal fragment AB$\overline{\text{B}}\overline{\text{L}}$," in *LICS*. IEEE Computer Society, 2011, pp. 387–396.

[7] V. Goranko, A. Montanari, and G. Sciavicco, "A road map of interval temporal logics and duration calculi," *Journal of Applied Non-Classical Logics*, vol. 14(1-2), pp. 9–54, 2004.

[8] J. Y. Halpern and Y. Shoham, "A propositional modal logic of time intervals," *Journal of the ACM*, vol. 38(4), pp. 935–962, 1991.

[9] M. R. Hansen, A. D. Phan, and A. W. Brekling, "A practical approach to model checking Duration Calculus using Presburger Arithmetic," *Annals of Mathematics and Artificial Intelligence*, vol. 71(1-3), pp. 251–278, 2014.

[10] K. Lodaya, "Sharpening the undecidability of interval temporal logic," in *ASIAN*, ser. LNCS 1961. Springer, 2000, pp. 290–298.

[11] A. Lomuscio and J. Michaliszyn, "An epistemic Halpern-Shoham logic," in *IJCAI*, 2013, pp. 1010–1016.

[12] ——, "Decidability of model checking multi-agent systems against a class of EHS specifications," in *ECAI*, 2014, pp. 543–548.

[13] J. Marcinkowski and J. Michaliszyn, "The undecidability of the logic of subintervals," *Fundamenta Informaticae*, vol. 131(2), pp. 217–240, 2014.

[14] A. Molinari, A. Montanari, A. Murano, G. Perelli, and A. Peron, "Checking Interval Properties of Computations," Dept. of Math and Computer Science, University of Udine, Tech. Rep. 2015/01, 2015, https://www.dimi.uniud.it/assets/preprints/1-2015-montanari.pdf.

[15] A. Molinari, A. Montanari, and A. Peron, "A Model Checking Procedure for Interval Temporal Logics based on Track Representatives," in *CSL*, 2015.

[16] A. Montanari, A. Murano, G. Perelli, and A. Peron., "Checking interval properties of computations," in *TIME*, 2014, pp. 59–68.

[17] A. Montanari, G. Puppis, and P. Sala, "Maximal decidable fragments of Halpern and Shoham's modal logic of intervals," in *ICALP (2)*, ser. LNCS 6199. Springer, 2010, pp. 345–356.

[18] B. Moszkowski, "Reasoning about digital circuits," Ph.D. dissertation, Dept. of Computer Science, Stanford University, Stanford, CA, 1983.

[19] R. J. Parikh, "On context-free languages," *Journal of the ACM*, vol. 13(4), pp. 570–581, 1966.

[20] I. Pratt-Hartmann, "Temporal prepositions and their logic," *Artificial Intelligence*, vol. 166(1-2), pp. 1–36, 2005.

[21] P. Roeper, "Intervals and tenses," *Journal of Philosophical Logic*, vol. 9, pp. 451–469, 1980.

[22] M. Sipser, *Introduction to the Theory of Computation*, 3rd ed. International Thomson Publishing, 2012.

[23] Y. Venema, "Expressiveness and completeness of an interval tense logic," *Notre Dame Journal of Formal Logic*, vol. 31(4), pp. 529–547, 1990.

[24] ——, "A modal logic for chopping intervals," *Journal of Logic and Computation*, vol. 1(4), pp. 453–476, 1991.

[25] C. Zhou and M. R. Hansen, *Duration Calculus - A Formal Approach to Real-Time Systems*, ser. Monographs in Theoretical Computer Science. An EATCS Series. Springer, 2004.